\documentclass[a4paper,11pt]{article}
\usepackage{pos}

\usepackage[subrefformat=parens]{subcaption}
\usepackage{xspace}
\usepackage[v1compatible,canvasScaleIsUnitLength]{axodraw2}

\makeatletter
\renewcommand\section{\@startsection {section}{1}{\z@}%
                                   {-2.5ex \@plus -1ex \@minus -.2ex}%
                                   {1.3ex \@plus.2ex}%
                                   {\normalfont\Large\bfseries}}
\makeatother

\newcommand{\bonsay}{\textsc{Bonsay}\xspace}
\newcommand{\recola}{\textsc{Recola}\xspace}
\newcommand{\openloops}{\textsc{OpenLoops2}\xspace}
\newcommand{\collier}{\textsc{Collier}\xspace}

\newcommand{\alphas}{\alpha_\mathrm{s}}

\newcommand{\ie}{\text{i.e.}\xspace}
\newcommand{\Pgne}{\nu_\mathrm{e}}
\newcommand{\Pgmp}{\mu^+}
\newcommand{\Pgngm}{\nu_\mu}
\newcommand{\Pqu}{\mathrm{u}}
\newcommand{\Pqc}{\mathrm{c}}
\newcommand{\Pqd}{\mathrm{d}}
\newcommand{\Pqs}{\mathrm{s}}

\def\mathswitchr#1{\relax\ifmmode{\mathrm{#1}}\else$\mathrm{#1}$\fi}

\newcommand{\PW}{\mathswitchr W}

\newcommand{\Pp}{\mathswitchr p}
\newcommand{\Pd}{\mathswitchr d}
\newcommand{\Pu}{\mathswitchr u}

\newcommand{\Pep}{\mathswitchr {e^+}}

\newcommand{\PWp}{\mathswitchr {W^+}}

\def\mathswitch#1{\relax\ifmmode#1\else$#1$\fi}

\newcommand{\MW}{\mathswitch {M_\PW}}

\newcommand{\lsim}
{\;\raisebox{-.3em}{$\stackrel{\displaystyle <}{\sim}$}\;}

\newcommand{\GeV}{\unskip\,\mathrm{GeV}}

\title{Full and approximated NLO predictions for like-sign W-boson scattering at the LHC}

\author*[a]{Stefan Dittmaier}
\author[b]{Christopher Schwan}
\author[c]{Ramon Winterhalder}

\affiliation[a]{Physikalisches Institut, Albert-Ludwigs-Universit\"at Freiburg,\\
  Hermann-Herder-Straße 3, 79104 Freiburg, Germany}

\affiliation[b]{Institut f\"ur Theoretische Physik und Astrophysik, Universit\"at W\"urzburg, \\
        Emil-Hilb-Weg 22, 97074 W\"urzburg, Germany}

\affiliation[c]{Centre for Cosmology, Particle Physics and Phenomenology (CP3),\\  Universit\'e catholique de Louvain,
        Chemin du Cyclotron 2, B-1348 Louvain-la-Neuve, Belgium}

\emailAdd{stefan.dittmaier@physik.uni-freiburg.de}
\emailAdd{christopher.schwan@physik.uni-wuerzburg.de}
\emailAdd{ramon.winterhalder@uclouvain.be}

\abstract{We report on a recent calculation of next-to-leading-order (NLO) QCD and electroweak
corrections to like-sign W-boson scattering at the Large Hadron Collider, including all
partonic channels and W-boson decays in the process
$\Pp \Pp \to \Pep \Pgne \, \Pgmp \Pgngm \, \mathrm{j} \mathrm{j} + X$.
The calculation is implemented in the Monte Carlo integrator \bonsay\ and 
comprises the full tower of NLO contributions of the orders 
$\alphas^3\alpha^4$, $\alphas^2\alpha^5$, $\alphas\alpha^6$, and $\alpha^7$.
Our numerical results confirm and extend previous results, in particular the occurrence
of large purely electroweak corrections of the order of $\sim-12\%$ for integrated cross sections,
which get even larger in distributions.
We construct a \textit{VBS approximation} for the NLO prediction based on 
partonic channels and gauge-invariant (sub)matrix elements potentially containing the 
vector-boson scattering (VBS) subprocess and on resonance expansions of the W~decays.
The VBS approximation reproduces the full NLO predictions within $\lsim1.5\%$
in the most important regions of phase space.
Moreover, we discuss results from different versions of 
\textit{effective vector-boson approximations}
at leading order, based on the collinear emission of W~bosons of incoming (anti)quarks.
However, owing to the only mild collinear enhancement and the design of VBS analysis cuts,
the quality of this approximation turns out to be only qualitative at the LHC.}

\FullConference{Loops and Legs in Quantum Field Theory (LL2024)\\
 14-19, April, 2024\\
Wittenberg, Germany\\}

\begin{document}
\maketitle

\section{Introduction}

Electroweak (EW) vector-boson scattering (VBS)
is among the most interesting classes of processes first accessible 
at the LHC to investigate the EW gauge structure and EW symmetry breaking.
Taking into account leptonic decays of the EW gauge bosons, the experimental
signature of VBS at the LHC is given by two mostly forward/backward-pointing
jets and four leptons, with at least two of them charged and the others being
neutrinos leading to missing transverse energy in the detector.
The corresponding $2\to6$ matrix elements do not only involve purely EW
diagrams, but also gluon exchange or even gluon-fusion channels,
leading to a large number of partonic channels and a whole tower of
next-to-leading order (NLO) corrections featuring not only pure QCD and EW
corrections but also QCD--EW mixed contributions.
Precision calculations for VBS processes have been continuously extended and refined
over the last few decades and culminated in the knowledge of the full towers
of NLO corrections for all relevant VBS processes at the LHC 
and the matching of fixed-order predictions with QCD parton showers
(see Ref.~\cite{Ballestrero:2018anz} 
and references therein).
The NLO corrections of all VBS processes share the feature that the
EW channels, whose relative contribution is enhanced by dedicated VBS selection cuts,
receive particularly large EW corrections of 10--15\%.

In this article we summarize the salient features and results of a recent
recalculation~\cite{Dittmaier:2023nac}
of the NLO corrections to like-sign WW scattering.
Previous results on QCD 
corrections~\cite{Jager:2009xx,Denner:2012dz}
including parton-shower matching~\cite{Jager:2011ms}
as well as on all NLO orders including pure EW and QCD--EW 
contributions~\cite{Biedermann:2016yds}
have been presented in the literature before.
Apart from providing cross-checks to existing results,
we also present an approximation of the NLO corrections
based on gauge-invariant subcontributions featuring the VBS subprocess
and on resonance expansions for the W-boson decays, as well as
a detailed discussion of the so-called \textit{effective vector-boson approximation}
for leading-order (LO) predictions
based on the picture of massive vector bosons as partons of the proton.

\section{Like-sign W-boson scattering at NLO}

\begin{figure}
\centering
\begin{subfigure}{0.33\textwidth}
\centering
\includegraphics{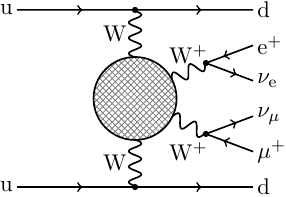}
\caption{VBS, doubly-resonant}
\label{fig:born_qq_vbs}
\end{subfigure}
\begin{subfigure}{0.33\textwidth}
\centering
\includegraphics{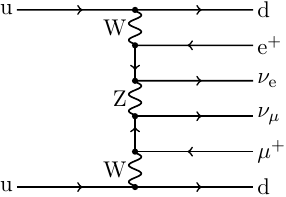}
\caption{non-resonant}
\label{fig:born_qq_nonres}
\end{subfigure}%
\begin{subfigure}{0.33\textwidth}
\centering
\includegraphics{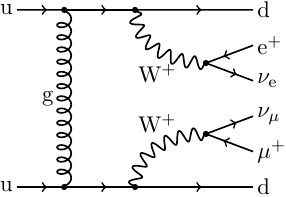}
\caption{QCD contribution}
\label{fig:born_qq_gluon}
\end{subfigure}%
\caption{Examples of LO Feynman diagrams 
for the partonic subprocess $\Pu \Pu \to \Pd \Pd \, \Pep \Pgne \, \Pgmp \Pgngm$.
The shaded blob represents tree-level subdiagrams for $\PWp \PWp\to \PWp \PWp$.}
\label{fig:born-diagrams}
\end{figure}

\begin{figure}
\centering
\begin{subfigure}{0.33\textwidth}
\centering
\includegraphics{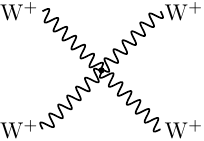}
\caption{4-point interaction}
\label{fig:vbs_4}
\end{subfigure}%
\begin{subfigure}{0.33\textwidth}
\centering
\includegraphics{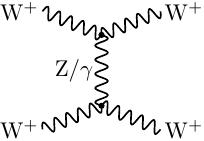}
\caption{$t$ channel}
\label{fig:vbs_t}
\end{subfigure}%
\begin{subfigure}{0.33\textwidth}
\centering
\includegraphics{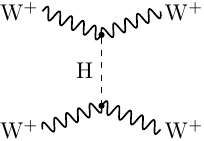}
\caption{$t$-channel Higgs}
\label{fig:vbs_t_higgs}
\end{subfigure}%
\caption{Typical VBS subdiagrams for $\PWp \PWp\to \PWp \PWp$ contained in the shaded blob of diagram Fig.~\ref{fig:born_qq_vbs}.}
\label{fig:vbs-diagrams}
\end{figure}

\begin{figure}
\centering
\begin{subfigure}{0.33\textwidth}
\centering
\includegraphics{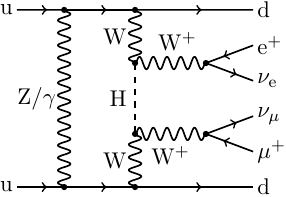}
\caption{6-point function with Higgs}
\label{fig:virt_Higgs}
\end{subfigure}%
\begin{subfigure}{0.33\textwidth}
\centering
\includegraphics{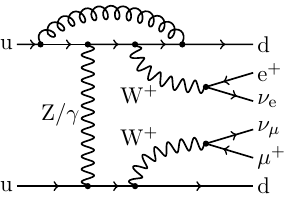}
\caption{gluon at single quark line}
\label{fig:virt_gl_sameline}
\end{subfigure}%
\begin{subfigure}{0.33\textwidth}
        \centering
        \includegraphics{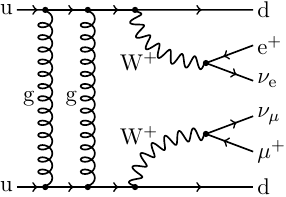}
        \caption{double gluon exchange}
        \label{fig:virt_gl_double_qcd}
\end{subfigure}%
\caption{Examples of one-loop diagrams for 
the partonic subprocess $\Pu \Pu \to \Pd \Pd \, \Pep \Pgne \, \Pgmp \Pgngm$.}
\label{fig:virtual-diagrams}
\end{figure}

Like-sign $\PW\PW$ production at the LHC has the signature of
two like-sign leptons, two jets, which are typically forward/backward pointing,
and missing transverse momentum carried away by two neutrinos.
For definiteness, we consider the process
$\Pp \Pp \to \Pep \Pgne \, \Pgmp \Pgngm \, \mathrm{j} \mathrm{j} + X$.
The LO cross section for $\PW^\pm\PW^\pm$ scattering
receives contributions from
$qq$, $q\bar q$, and $\bar q\bar q$ partonic channels involving
various (anti)quark combinations with positive/negative net electric charge,
i.e.\ at LO there are no partonic channels {involving initial-state gluons.}
Denoting the EW and the strong coupling constants as $e$ and $g_\mathrm{s}$, 
respectively, the LO matrix elements scale like $\mathcal{O}(e^6)$
or $\mathcal{O}(g_\mathrm{s}^2 e^4)$;
the respective contributions to the LO matrix element are denoted
$\mathcal{M}_{e^6}$ and $\mathcal{M}_{g_\mathrm{s}^2 e^4}$ in the following.
The structural diagram featuring the VBS subprocess,
an EW background diagram as well as a diagram with gluon exchange
are illustrated in Fig.~\ref{fig:born-diagrams}.
The different types of VBS subdiagrams contained in $\mathcal{M}_{e^6}$
are shown in Fig.~\ref{fig:vbs-diagrams}.
Squaring the LO matrix element leads to three different perturbative orders,
scaling like $\alpha^6$, $\alpha_\mathrm{s}^2 \alpha^4$, and $\alpha_\mathrm{s} \alpha^5$.
The latter interference contribution is numerically strongly suppressed due to its
colour structure (which demands identical quark generations in the two quark chains)
and due to the fact that forward- or backward-enhanced $t/u$-channel propagators
are not further enhanced by squaring.
At NLO, each of the LO matrix elements can receive EW and QCD corrections, i.e.\
there are {one-loop amplitudes of $\mathcal{O}(e^8)$,
$\mathcal{O}(g_\mathrm{s}^2 e^6)$, and $\mathcal{O}(g_\mathrm{s}^4 e^4)$ at NLO.}
Some corresponding  one-loop diagrams of these orders are shown in
Fig.~\ref{fig:virtual-diagrams}.

The cross sections are evaluated with
the Monte Carlo program \bonsay, which is based on multi-channel Monte Carlo integration using
adaptive weight 
optimization~\cite{Berends:1994xn},
similar to the approach described in Ref.~\cite{Dittmaier:2002ap}.
\bonsay supports the parallel computation of uncertainties induced by different
scale choices and errors in parton distribution functions.
Both the tree-level and one-loop matrix elements are provided by
\openloops~\cite{Cascioli:2011va} 
by default and
have been cross-validated against respective results obtained with
\recola~\cite{Actis:2016mpe}.
The one-loop integrals are numerically evaluated using
\collier~\cite{Denner:2016kdg}, which employs the methods and results of 
Ref.~\cite{Denner:2002ii}
to numerically stabilize the results 
in the vicinity of exceptional phase-space configurations.
Particle resonances are described in the complex-mass
scheme~\cite{Denner:1999gp,Denner:2019vbn}
to guarantee gauge independence of amplitudes and NLO accuracy 
in both resonant and non-resonant phase-space regions.
The extraction and cancellation of (soft and collinear) infrared singularities 
is accomplished within the dipole subtraction formalism both for
QCD~\cite{Catani:1996vz} and EW~\cite{Dittmaier:1999mb} 
corrections.

In addition to performing the full NLO calculation we have constructed 
a {\textit{VBS approximation (VBSA)}} which can serve as a proxy for the full
calculation with a precision that is sufficient for most phenomenological analyses
and which is much less costly in terms of CPU time.
At LO, the VBSA keeps the full $2\to6$ matrix elements.
At NLO, the VBSA merges two different approximative steps:
Similar to the approach used in Ref.~\cite{Denner:2012dz},
step~1 selects all partonic channels that contain the VBS subprocess and further
strips contributions of minor importance. In detail, matrix elements are
decomposed into gauge-invariant parts characterized by different
fermion-number flows, and only channels related to VBS are kept.
This step, in particular, eliminates all channels featuring WWW production instead of VBS.
Moreover, in the process of squaring the amplitude, all interference terms from
different fermion-number flows are discarded, since they are colour suppressed and do not
receive the kinematical VBS enhancements from squared $t/u$-channel W~propagators.
In step~1, thus, all one-loop amplitudes can be constructed from one
prototype channel, e.g.\ $\Pqc \Pqu \to \Pep \Pgne \, \Pgmp \Pgngm \, \Pqd \Pqs$,
via crossing.
Step~2 in the VBSA construction applies the \textit{double-pole approximation (DPA)}
to the produced W~bosons in the virtual corrections, i.e.\ all one-loop
matrix elements are expanded about the two W~resonance poles. This procedure splits the virtual
corrections into \textit{factorizable} and \textit{non-factorizable} parts, the former
containing the corrections to the W-pair production and the W-decay subprocesses,
the latter accounting for doubly-resonant effects from soft-photon exchange between
the subprocesses. More details on the DPA concept and the specifically employed
variant can be found in
Refs.~\cite{Denner:2000bj,Dittmaier:2015bfe,Denner:2019vbn}
and references therein.
We finally note that care has to be taken in the approximation of the real-emission
corrections since they are related to different underlying LO channels in different
collinear limits of phase space. 
For the full construction of the VBSA we refer to 
Ref.~\cite{Dittmaier:2023nac}.

\section{Cross-section predictions at NLO}

\begin{figure}
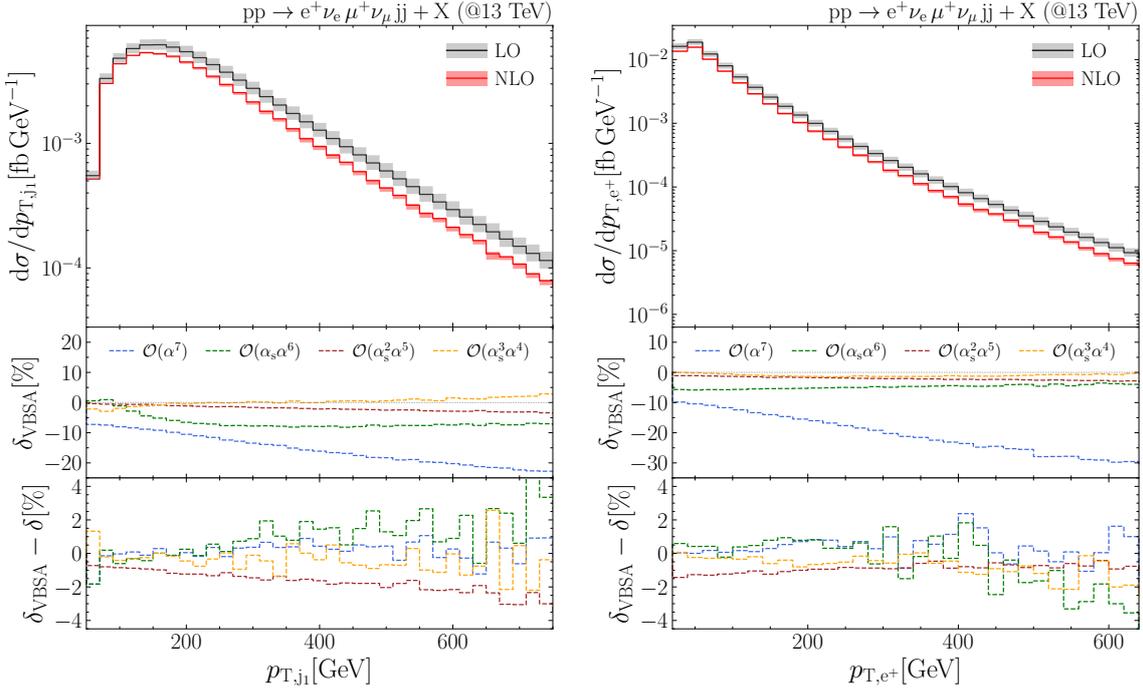

\includegraphics[bb= 10 3 458 557,clip,page=9,width=0.49\textwidth]{nlos_vs_vbsa} \hfill
\includegraphics[bb= 10 3 458 557,clip,page=8,width=0.49\textwidth]{nlos_vs_vbsa}
\caption{Distributions in the transverse momenta of the leading jet and the positron
and corresponding corrections:
absolute predictions (top panels), relative NLO corrections in VBSA (middle panels),
and difference between relative full NLO corrections and corresponding VBSA (bottom panels).
[Taken from Ref.~\cite{Dittmaier:2023nac}.]}
\label{fig:NLO-pTj-pTl}
\end{figure}

\begin{figure}
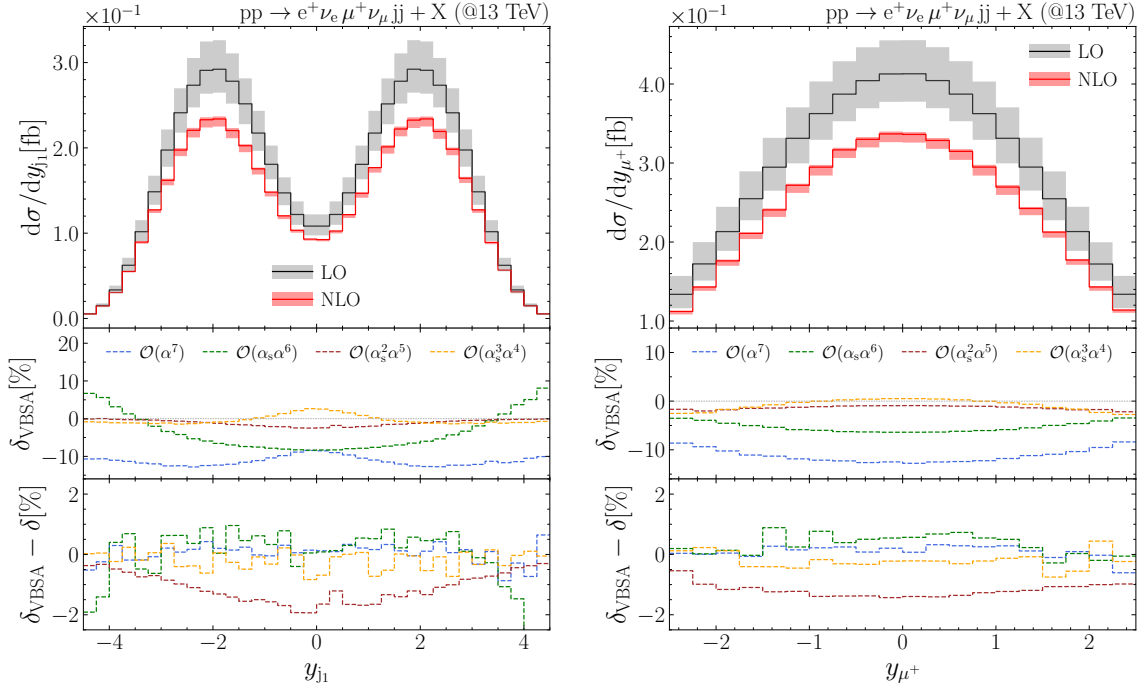

\includegraphics[bb= 10 3 458 557,clip,page=3,width=0.49\textwidth]{nlos_vs_vbsa} \hfill
\includegraphics[bb= 10 3 458 557,clip,page=5,width=0.49\textwidth]{nlos_vs_vbsa}
\caption{As in Fig.~\ref{fig:NLO-pTj-pTl}, but for the distributions in the
rapidities of the leading jet and the muon.
[Taken from Ref.~\cite{Dittmaier:2023nac}.]}
\label{fig:NLO-etaj-etal}
\end{figure}

\begin{figure}
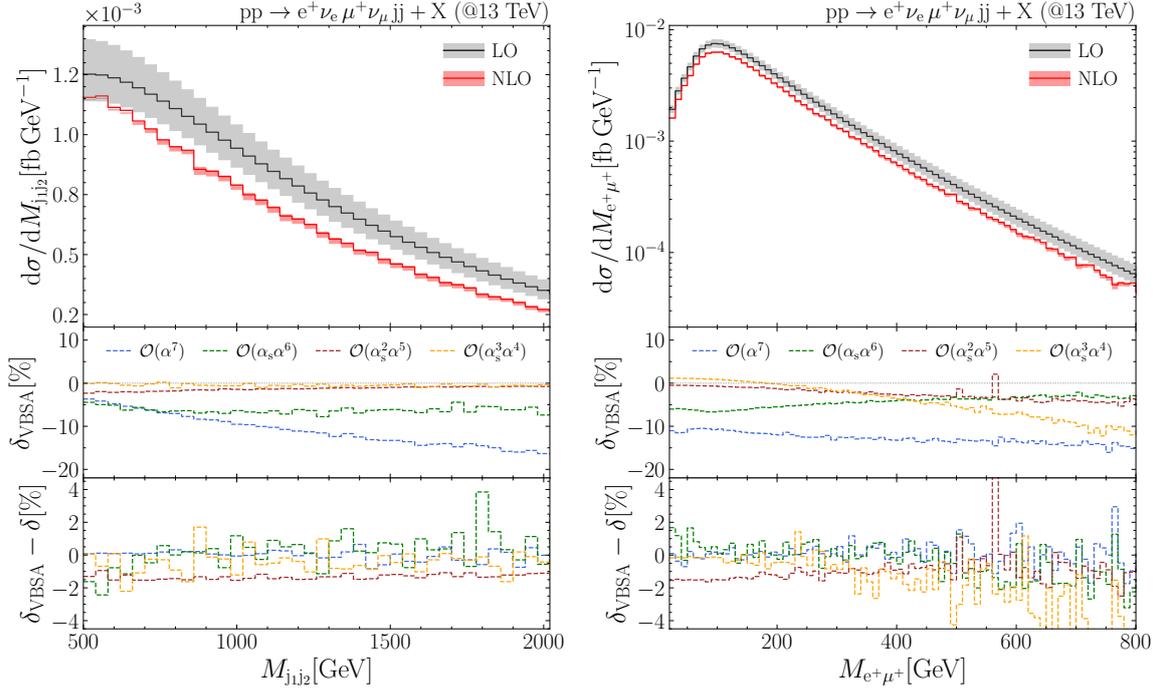

\includegraphics[bb= 10 3 458 557,clip,page=7,width=0.49\textwidth]{nlos_vs_vbsa} \hfill
\includegraphics[bb= 10 3 458 557,clip,page=6,width=0.49\textwidth]{nlos_vs_vbsa}
\caption{As in Fig.~\ref{fig:NLO-pTj-pTl}, but for the
distributions in the jet-pair (left) and charged-lepton-pair (right).
[Taken from Ref.~\cite{Dittmaier:2023nac}.]}
\label{fig:NLO-Mjj-Mll}
\end{figure}

In Figs.~\ref{fig:NLO-pTj-pTl}--\ref{fig:NLO-Mjj-Mll}
we compare the full relative NLO corrections
$\delta = \mathrm{d}\sigma_\text{NLO}/\mathrm{d}\sigma_\text{LO}-1$
with its approximated version
{$\delta_\text{VBSA} = \mathrm{d}\sigma_\text{NLO, VBSA}/\mathrm{d}\sigma_\text{LO}-1$
in VBSA.
The bands in the figures illustrate the residual scale uncertainties,
obtained from the envelope of the usual seven-point scale variation with factors
of $0.5$ and $2$ using the 
geometric mean of the transverse momenta of the tagging jets as central scale.}

Figure~\ref{fig:NLO-pTj-pTl} shows the results for  the distributions in the
transverse momenta of the leading jet and the positron.
In both cases, for moderate and large $p_{\mathrm{T}}$, the purely EW corrections of
$\mathcal{O}(\alpha^7)$ are dominant, negative, and increasing in magnitude for
increasing transverse momenta, reaching about $\sim-20\%$ and $\sim-30\%$
at $p_{\mathrm{T}}=600\GeV$, respectively.
This behaviour is a typical sign for the appearance of EW Sudakov logarithms
at high energies.
The different size of the {effects} for jets and W-decay leptons can be understood as follows:
The $p_{\mathrm{T}}$ distributions of the jets 
do not entirely zoom into the Sudakov regime
of the WW$\to$WW subprocess, which demands large Mandelstam variables in the $2\to2$ subprocess and small virtualities of the incoming (off-shell) W~bosons.
For large $p_{\mathrm{T}}$ values of a jet, the virtuality of at least one of the incoming W~bosons is not small, and the $t$-channel momentum transfer in the WW$\to$WW subprocess is not forced to be large.
Therefore, the EW Sudakov double logarithms cannot fully dominate the corrections to the
$p_{\mathrm{T}}$ of the jets, and all kinds of nominally {subleading}
EW high-energy corrections become relevant.
On the other hand, the domain of large $p_{\mathrm{T}}$ of any of the decay leptons is
dominated by the Sudakov regime of the WW$\to$WW subprocess,
because the preference of small jet transverse momenta leads to small
virtualities of the incoming W~bosons and the large transverse momentum of a decay
lepton requires both a large scattering energy (Mandelstam variable $s$)
and large momentum transfer (Mandelstam variable $t$) of the subprocess.
The observed $\sim-30\%$ can be qualitatively reproduced by just taking into account
the EW Sudakov correction factor for the WW$\to$WW subprocess,
$\delta_{\mathrm{Sud}}=-\frac{2\alpha}{\pi s_{\mathrm{w}}^2}\ln^2(\hat s/\MW^2)$,
where $\sqrt{\hat s}=\mathcal{O}(p_{\mathrm{T}})$ is the WW centre-of-mass energy and
$s_{\mathrm{w}}$ the sine of the weak mixing angle.
These features were already highlighted in
Ref.~\cite{Biedermann:2016yds}.
The next-to-largest corrections to the $p_{\mathrm{T}}$ distributions 
are the mixed QCD--EW corrections of $\mathcal{O}(\alpha_s\alpha^6)$,
which amount to 5--10\% above the maximum in the leading-jet distribution
and are almost uniformly $\sim -5\%$ in the $p_{\mathrm{T}}$ spectrum of the electron.
The suppression of the remaining corrections of $\mathcal{O}(\alpha_s^2\alpha^5)$
and $\mathcal{O}(\alpha_s^3\alpha^4)$ is mostly due to the fact that the
cross-section contributions widely inherit the kinematic behaviour of the
LO QCD contribution, which is small compared to the EW contribution over the
whole distribution as a consequence of the VBS cuts.
The approximative quality of the VBSA is typically at the 1\%~level 
for all orders of the NLO tower 
in the regions of phase space where the relevant part of the cross section is 
concentrated,
with the exception of the $\mathcal{O}(\alpha_s^2\alpha^5)$ contribution
where the difference $\delta_\text{VBSA}-\delta$ can reach the order of 1.5\% in
size.

Figure~\ref{fig:NLO-etaj-etal} shows the distributions in the
rapidities of the leading jet and the muon.
The hierarchy among the various NLO contributions is similar as for the $p_{\mathrm{T}}$
distributions shown above, \ie the purely EW corrections of $\mathcal{O}(\alpha^7)$ are
the dominating ones, followed by the order $\mathcal{O}(\alpha_s\alpha^6)$, while
the other two orders with higher powers of $\alpha_s$ are widely suppressed.
The corrections of $\mathcal{O}(\alpha^7)$ show much less variations in shape
than for the $p_{\mathrm{T}}$ distributions and are typically about $-10\%$ to
$-12\%$. This is due to the fact that the large logarithmic EW high-energy
corrections uniformly contribute to all rapidities, in contrast
to the $p_{\mathrm{T}}$ distributions where they appear at high scales only.
The moderate variations in the $\mathcal{O}(\alpha^7)$ corrections mostly result
from the change in the LO normalization induced by the variation in its
composition from EW and QCD parts; normalizing the $\mathcal{O}(\alpha^7)$
contribution to the $\mathcal{O}(\alpha^6)$ LO part would produce a nearly flat relative
$\mathcal{O}(\alpha)$ correction.
The overall second-largest corrections are again the ones of
$\mathcal{O}(\alpha_s\alpha^6)$, which are dominated by the QCD corrections to the
EW LO channel. Their largest impact, growing even to $\sim10\%$, is on the leading jet
at high rapidities.
In the other rapidity regions those corrections hardly exceed 5\%.
The pure QCD corrections of $\mathcal{O}(\alpha_s^3\alpha^4)$ only exceed the 1\% level
significantly for central rapidities.
The mixed corrections of order $\mathcal{O}(\alpha_s^2\alpha^5)$ never exceed the
1\% level at all.
Whenever the cross section is sizeable,
the approximative quality of the VBSA is again at the level of 1\% or better for
all NLO orders but $\mathcal{O}(\alpha_s^2\alpha^5)$, where it is of the order of
1.5\%.

{Finally, we show the distributions in the invariant mass of the
two leading jets (left) and in} the invariant mass of the charged leptons (right) in 
Fig.~\ref{fig:NLO-Mjj-Mll}.
The hierarchy of the various corrections and their behaviour can be widely explained following
similar arguments as above.
The trend of the dominating genuine weak corrections of $\mathcal{O}(\alpha^7)$ towards
increasingly negative corrections for larger scales is visible as for the
$p_{\mathrm{T}}$ distributions, but the increase in size to about $-15\%$ for the largest
considered scales is much less dramatic. This is due to the fact that the domain
of large invariant masses $M_{\mathrm{j}_1\mathrm{j}_2}$ or $M_{\Pep\mu^+}$
is not fully dominated by the Sudakov regime of the WW$\to$WW subprocess,
because the $t/u$-channel-like momentum transfer in the subprocess is not forced to be large.
Thus, the impact of the leading EW Sudakov corrections is
damped to the size of the {subleading} EW high-energy corrections.
In the $M_{\mathrm{j}_1\mathrm{j}_2}$ distribution, which is mostly dominated by
EW contributions at LO,
the corrections of $\mathcal{O}(\alpha_s\alpha^6)$ typically have an impact at the
5\%~level, while the remaining two NLO orders with higher powers of $\alpha_s$ hardly
reach 1\%.
The mixed QCD--EW and the pure QCD corrections show, however, an interesting crossover
in the $M_{\Pep\mu^+}$ distribution at $M_{\Pep\mu^+}\sim400\GeV$, which we attribute
to the increasing influence of the LO QCD contribution.
For $M_{\Pep\mu^+}<400\GeV$, where the EW part strongly dominates the LO cross section,
the $\mathcal{O}(\alpha_s\alpha^6)$ correction is the second largest
after the genuine EW correction, and the remaining NLO orders are at the 1\%~level.
For $M_{\Pep\mu^+}>400\GeV$, where the LO QCD part competes in size with the EW LO part,
the corrections of
$\mathcal{O}(\alpha_s^2\alpha^5)$ and $\mathcal{O}(\alpha_s^3\alpha^4)$ dominate over
$\mathcal{O}(\alpha_s\alpha^6)$ and reach $\sim-5\%$ for large $M_{\Pep\mu^+}$.
Similar to the previously considered distributions,
the approximative quality of the VBSA is at the level of 1.5\% for the
$\mathcal{O}(\alpha_s^2\alpha^5)$ corrections and of 1\% for the other
NLO orders.

\section{Effective W-boson approximation at LO}

The \textit{effective vector-boson approximation (EVA),} the idea of which goes back to 
Ref.~\cite{Dawson:1984gx},
extends the idea of partons inside hadrons to the case of weak vector bosons,
which play the role of partons in (anti)quarks,
just like (anti)quarks and gluons in hadrons.
The vector-boson emission $q\to qV$ is approximated by its asymptotic behaviour 
in the collinear limit, where it is logarithmically enhanced. 
Previous studies in the literature have already indicated 
that the approximation quality of the EVA is rather limited
(see, e.g., Refs.~\cite{Kuss:1995yv,Accomando:2006mc}
and references therein).

A comprehensive description of our construction of the LO EVA matrix elements is
given in the appendix of Ref.~\cite{Dittmaier:2023nac}.
Figure~\ref{fig:EVBAdiag} schematically illustrates the factorization of VBS 
matrix elements into
W~radiation off (anti)quarks, VBS core process, and subsequent W-boson decays.
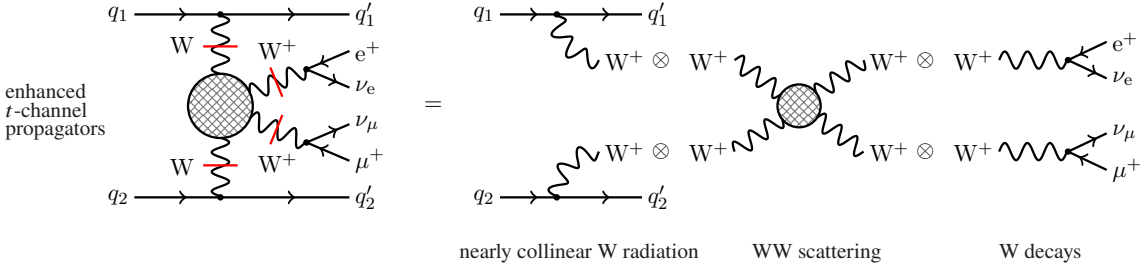
\begin{figure}
\begin{picture}(0,0)(-15,0)
\LongArrow(25,40)(47,50)
\LongArrow(25,35)(47,25)
\put(-15, 42){\scriptsize enhanced}
\put(-15, 35){\scriptsize $t$-channel}
\put(-15, 28){\scriptsize propagators}
\end{picture}
\hspace*{3em}
\includegraphics[width=.9\textwidth]{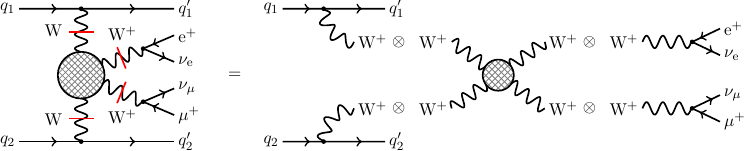}%
\\[.3em]
{\scriptsize
\hspace*{21em} nearly collinear $\PW$ radiation
\hspace*{2em} $\PW\PW$ scattering
\hspace*{5em} \rlap{$\PW$ decays}
}
\caption{Schematic illustration of the EVA factorization of VBS matrix elements.}
\label{fig:EVBAdiag}
\end{figure}
We do not merely take over
existing proposals from the literature, but compare various formulations that differ
in the details of handling intermediate (off-shell) polarization vectors and
external currents describing the W~radiation off the (anti)quarks and the W~decays
into leptons, in order to account for spin correlations and off-shell effects
as much as possible.

\begin{figure}
\includegraphics[page=8,width=0.49\textwidth]{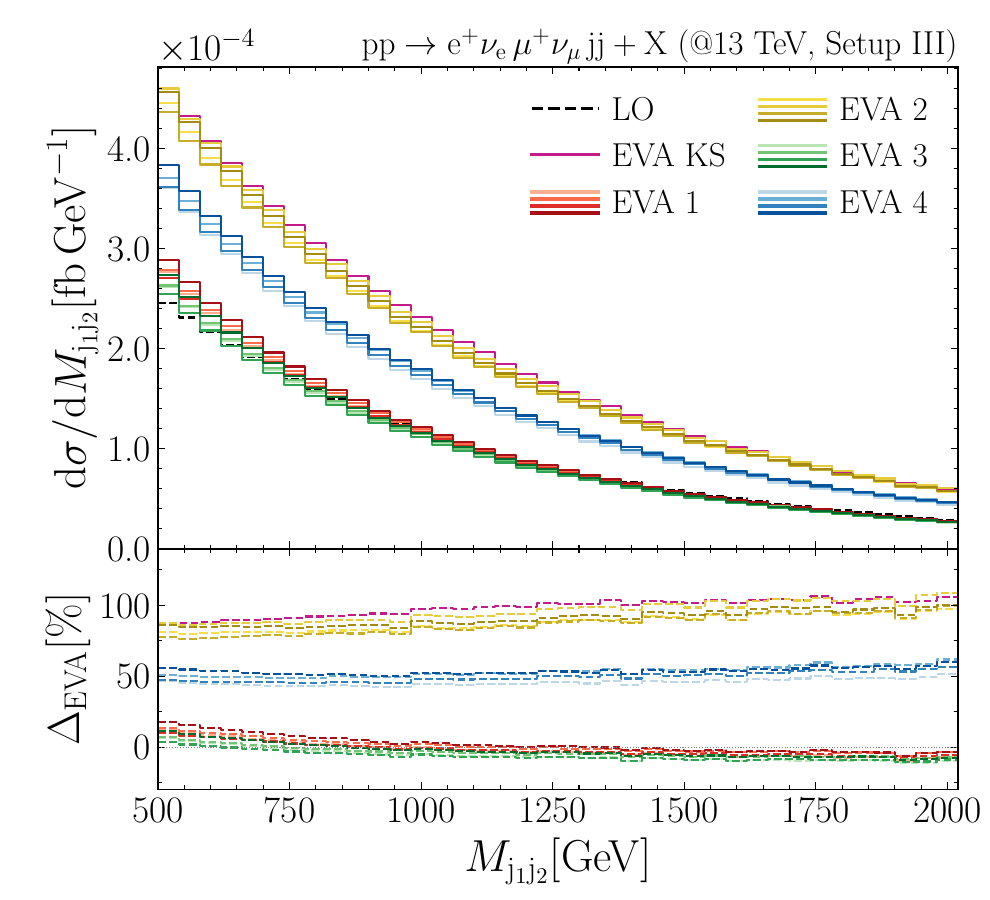} \hfill
\includegraphics[page=7,width=0.49\textwidth]{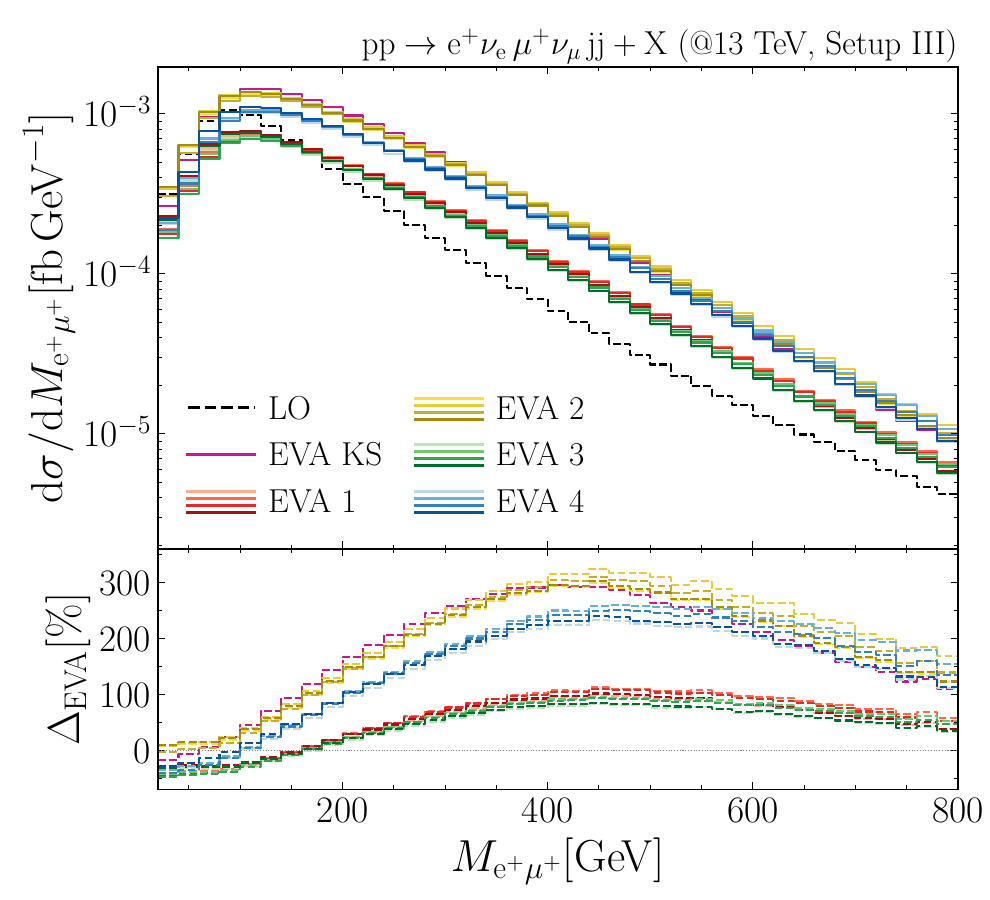}
\caption{LO distributions in the invariant masses of the two jets and the two
charged leptons, respectively, in like-sign WW scattering,
based on full LO matrix elements (LO) and different EVA versions (EVA1--4 and EVA~KS,
see main text), employing typical VBS selection cuts but with an ``inverted
cut'' on the jets transverse momenta, $p_{\mathrm{T,jet}}<100\GeV$.
The relative deviation of the EVA predictions from LO
is quantified by $\Delta_{\mathrm{EVA}}
=\mathrm{d}\sigma_{\mathrm{EVA}}/\mathrm{d}\sigma_{\mathrm{LO}}-1$. 
[Taken from Ref.~\cite{Dittmaier:2023nac}.]}
\label{fig:EVBAdistributions}
\end{figure}

Figure~\ref{fig:EVBAdistributions} exemplarily illustrates the quality of different
EVA versions for the invariant-mass distributions of the two jets and the two
charged leptons, respectively.
More results can be found in Ref.~\cite{Dittmaier:2023nac}.
The most important difference between the EVA versions is that
EVA1 and EVA3 restore the transversality of the polarization vectors
of initial-state W~bosons in VBS amplitude, but not EVA2 and EVA4. 
Other, less important differences concern 
the transversality conditions for the final-state W~bosons of the VBS process
and for the leptonic decay currents, and an optional 
relative sign factor between transverse and longitudinal polarization vectors 
of the incoming (off-shell) W~bosons, as described in detail in Ref.~\cite{Dittmaier:2023nac}.
EVA~KS refers to the EVA variant of 
Ref.~\cite{Kuss:1995yv},
where an extra weight factor was introduced for longitudinal incoming
off-shell W~bosons.
All EVA versions are evaluated with an on-shell projection of the W~momenta in the
VBS subprocess that forces the W~momenta on-shell to guarantee gauge independence
{of the WW$\to$WW matrix elements}
and preserves the locations of the photon
poles in the $t$- and $u$-channel subdiagrams~(b) of Fig.~\ref{fig:vbs-diagrams}
to avoid extra enhancements in the intrinsic uncertainty of the EVA.

{Generically,} we find that all EVA versions can only qualitatively describe
the full VBS process as long as typical VBS selection cuts are applied, which
exclude the very forward region of jet emission for which the EVA is actually 
designed. Only if we invert the cut on the jets transverse momenta to
$p_{\mathrm{T,jet}}<100{-}150\GeV$, the EVA delivers results of some reasonable
approximative quality. In regions where the cross section is maximal,
some EVA versions are good within 10--20\% for distributions defined from the jet
kinematics, but none are better than 50--100\% for leptonic observables.

\section{Conclusions}

We have reported on a recent calculation~\cite{Dittmaier:2023nac}
of the full tower of NLO 
corrections to like-sign W-boson scattering at the LHC, including all
partonic channels and W-boson decays.
Our calculation, which is implemented in the Monte Carlo integrator \bonsay,
confirms the results of a previous calculation 
(up to a glitch in a numerically unimportant contribution)
and in particular the occurrence
of large pure EW corrections of the order of $\sim-12\%$ for integrated cross sections.

Moreover, we have constructed a \textit{VBS approximation} for the NLO prediction based on 
partonic channels and gauge-invariant (sub)matrix elements featuring the 
VBS subprocess and on resonance expansions of the W~decays.
The VBS approximation reproduces the full NLO predictions within $\lsim1.5\%$
in the most important regions of phase space.
Finally, we have discussed results from different versions of 
\textit{effective vector-boson approximations}
at LO, based on the collinear emission of W~bosons of incoming (anti)quarks.
In line with previous findings for similarly constructed approximations,
we find that the approximative quality is only qualitative at the LHC
owing to the mild collinear enhancement of the W-boson emission and 
the design of VBS analysis cuts, which excludes very forward/backward pointing jets.


\begin{thebibliography}{99}

\bibitem{Ballestrero:2018anz}
A.~Ballestrero, \textit{et al.}
Eur. Phys. J. C \textbf{78} (2018) no.8, 671
[arXiv:1803.07943 [hep-ph]];\\
%
R.~Covarelli, M.~Pellen and M.~Zaro,
Int. J. Mod. Phys. A \textbf{36} (2021) no.16, 2130009
[arXiv:2102.10991 [hep-ph]];\\
%
D.~Buarque Franzosi, \textit{et al.}
Rev. Phys. \textbf{8} (2022), 100071
[arXiv:2106.01393 [hep-ph]].

\bibitem{Dittmaier:2023nac}
S.~Dittmaier, \textit{et al.}
JHEP \textbf{11} (2023), 022
[arXiv:2308.16716 [hep-ph]].

\bibitem{Jager:2009xx}
B.~{J\"ager}, C.~Oleari and D.~Zeppenfeld,
Phys. Rev. D \textbf{80} (2009), 034022
[arXiv:0907.0580 [hep-ph]];\\
%
T.~Melia, \textit{et al.}
JHEP \textbf{12} (2010), 053
[arXiv:1007.5313 [hep-ph]];\\
%
F.~Campanario, \textit{et al.}
Phys. Rev. D \textbf{89} (2014) no.5, 054009
[arXiv:1311.6738 [hep-ph]].

\bibitem{Denner:2012dz}
A.~Denner, L.~Hosekova and S.~Kallweit,
Phys. Rev. D \textbf{86} (2012), 114014
[arXiv:1209.2389 [hep-ph]].

\bibitem{Jager:2011ms}
B.~{J\"ager} and G.~Zanderighi,
JHEP \textbf{11} (2011), 055
[arXiv:1108.0864 [hep-ph]].

\bibitem{Biedermann:2016yds}
B.~Biedermann, A.~Denner and M.~Pellen,
Phys. Rev. Lett. \textbf{118} (2017) no.26, 261801
[arXiv:1611.02951 [hep-ph]]; 
%
JHEP \textbf{10} (2017), 124
[arXiv:1708.00268 [hep-ph]];\\
%
M.~Chiesa, \textit{et al.}
Eur. Phys. J. C \textbf{79} (2019) no.9, 788
[arXiv:1906.01863 [hep-ph]].

\bibitem{Berends:1994xn}
F.~A.~Berends, R.~Pittau and R.~Kleiss,
Comput. Phys. Commun. \textbf{85} (1995), 437-452
[arXiv:hep-ph/9409326];
%
Comput. Phys. Commun. \textbf{85} (1995), 437-452
[arXiv:hep-ph/9409326];\\
%
R.~Kleiss and R.~Pittau,
Comput. Phys. Commun. \textbf{83} (1994), 141-146
[arXiv:hep-ph/9405257].

\bibitem{Dittmaier:2002ap}
S.~Dittmaier and M.~Roth,
Nucl. Phys. B \textbf{642} (2002), 307-343
[arXiv:hep-ph/0206070].

\bibitem{Cascioli:2011va}
F.~Cascioli, P.~{Maierh\"ofer} and S.~Pozzorini,
Phys. Rev. Lett. \textbf{108} (2012), 111601
[arXiv:1111.5206 [hep-ph]];\\
%
F.~Buccioni, \textit{et al.}
Eur. Phys. J. C \textbf{79} (2019) no.10, 866
[arXiv:1907.13071 [hep-ph]].

\bibitem{Actis:2016mpe}
S.~Actis, \textit{et al.}
Comput. Phys. Commun. \textbf{214} (2017), 140-173
[arXiv:1605.01090 [hep-ph]]; \\
%
A.~Denner, J.~N.~Lang and S.~Uccirati,
Comput. Phys. Commun. \textbf{224} (2018), 346-361
[arXiv:1711.07388 [hep-ph]].

\bibitem{Denner:2016kdg}
A.~Denner, S.~Dittmaier and L.~Hofer,
Comput. Phys. Commun. \textbf{212} (2017), 220-238
[arXiv:1604.06792 [hep-ph]].

\bibitem{Denner:2002ii}
A.~Denner and S.~Dittmaier,
Nucl. Phys. B \textbf{658} (2003), 175-202
[arXiv:hep-ph/0212259]; 
%
Nucl. Phys. B \textbf{734} (2006), 62-115
[arXiv:hep-ph/0509141]; 
%
Nucl. Phys. B \textbf{844} (2011), 199-242
[arXiv:1005.2076 [hep-ph]].

\bibitem{Denner:1999gp}
A.~Denner, \textit{et al.}
Nucl. Phys. B \textbf{560} (1999), 33-65
[arXiv:hep-ph/9904472];\\
%
A.~Denner, \textit{et al.}
Nucl. Phys. B \textbf{724} (2005), 247-294
[erratum: Nucl. Phys. B \textbf{854} (2012), 504-507]
[arXiv:hep-ph/0505042].

\bibitem{Denner:2019vbn}
A.~Denner and S.~Dittmaier,
Phys. Rept. \textbf{864} (2020), 1-163
[arXiv:1912.06823 [hep-ph]].

\bibitem{Catani:1996vz}
S.~Catani and M.~H.~Seymour,
Nucl. Phys. B \textbf{485} (1997), 291-419
[erratum: Nucl. Phys. B \textbf{510} (1998), 503-504]
[arXiv:hep-ph/9605323];\\
%
S.~Catani, \textit{et al.}
Nucl. Phys. B \textbf{627} (2002), 189-265
[arXiv:hep-ph/0201036].

\bibitem{Dittmaier:1999mb}
S.~Dittmaier,
Nucl. Phys. B \textbf{565} (2000), 69-122
[arXiv:hep-ph/9904440];\\
%
S.~Dittmaier, A.~Kabelschacht and T.~Kasprzik,
Nucl. Phys. B \textbf{800} (2008), 146-189
[arXiv:0802.1405 [hep-ph]].

\bibitem{Denner:2000bj}
A.~Denner, \textit{et al.}
Nucl. Phys. B \textbf{587} (2000), 67-117
[arXiv:hep-ph/0006307].

\bibitem{Dittmaier:2015bfe}
S.~Dittmaier and C.~Schwan,
Eur. Phys. J. C \textbf{76} (2016) no.3, 144
[arXiv:1511.01698 [hep-ph]].

\bibitem{Dawson:1984gx}
S.~Dawson,
Nucl. Phys. B \textbf{249} (1985), 42-60;\\
%
M.~S.~Chanowitz and M.~K.~Gaillard,
Phys. Lett. B \textbf{142} (1984), 85-90;\\
%
G.~L.~Kane, W.~W.~Repko and W.~B.~Rolnick,
Phys. Lett. B \textbf{148} (1984), 367-372;\\
%
J.~Lindfors,
Z. Phys. C \textbf{28} (1985), 427.

\bibitem{Kuss:1995yv}
I.~Kuss and H.~Spiesberger,
Phys. Rev. D \textbf{53} (1996), 6078-6093
[arXiv:hep-ph/9507204];\\
%
I.~Kuss,
Phys. Rev. D \textbf{55} (1997), 7165-7182
[arXiv:hep-ph/9608453].

\bibitem{Accomando:2006mc}
E.~Accomando, \textit{et al.}
Phys. Rev. D \textbf{74} (2006), 073010
[arXiv:hep-ph/0608019];\\
%
W.~Bernreuther and L.~Chen,
Phys. Rev. D \textbf{93} (2016) no.5, 053018
[arXiv:1511.07706 [hep-ph]].


\end{thebibliography}

\setlength{\bibsep}{0pt plus 0.3ex}

\end{document}